# *Ab initio* Investigation of Vibrational, Thermodynamic, and Optical properties of Sc$_2$AlC MAX compound

M. A. Ali[a], M. T. Nasir[d], M. R. Khatun[b], A. K. M. A. Islam[c], S. H. Naqib[b*]

[a]Department of Physics, Chittagong University of Engineering and Technology, Chittagong-4349, Bangladesh.
[b]Department of Physics, University of Rajshahi, Rajshahi-6205, Bangladesh.
[c]International Islamic University Chittagong, 154/A College Road, Chittagong, Bangladesh.
[d]Department of Arts &Science, Bangladesh Army University of Science and Technology, Saidpur-5310, Nilphamari, Bangladesh

**A B S T R A C T**

The structural, vibrational, thermodynamical and optical properties of technologically important, weakly coupled MAX compound, Sc$_2$AlC are calculated using density functional theory (DFT). The structural properties of Sc$_2$AlC are compared with results reported earlier. The vibrational, thermodynamical, and optical properties are theoretically estimated for the first time. The phonon dispersion curve has been calculated and the dynamical stability of this compound has been investigated. The optical and acoustic modes are observed clearly. We have calculated the Helmholtz free energy ($F$), internal energy ($E$), entropy ($S$) and specific heat capacity ($C_v$) from the phonon density of states. Various optical parameters have also been calculated. The reflectance spectrum shows that it this compound has the potential to be used as a solar reflector.

*Keywords:* MAX compound; Phonon dispersion; Thermodynamical properties; Optical properties

## 1. Introduction

Sc$_2$AlC belongs to the prototype of a vast family of ternary nitrides and carbides widely known as MAX compounds [1]. These materials represent a class of condensed phases that can be regarded as thermodynamically stable nanolaminates. MAX phases have attracted significant attention of the scientific community because of their striking combination of properties, some of which are like ceramics and the others metallic [2].

*Corresponding author. Email: salehnaqib@yahoo.com

To be specific, these compounds possess machinability, damage tolarence, thermal and electrical conductivity like metals with low density. On the other hand, they possess high elastic stiffness, refractory nature and are resistant to high-temperature oxidation, just like ceramics [3]. All these attributes make MAX phases industrially important materials for high performance applications in diverse fields from defense materials to electronic devices. So far, over 70 different MAX phases have been synthesized [4], $Sc_2AlC$ is one of them. The $M_2AX$ phases with M = Ti, V, Cr, Nb, Ta, Zr, Hf; A= Al, S, Sn, As, In, Ga, and X = N, C, have been studied extensively both experimentally and theoretically [4]. Nevertheless, the $Sc_2AlC$ phase remains one of the least studied materials among 211 MAX family.

A few first principles calculations of $Sc_2AlC$ phase can be found in the literature. The structural and elastic properties have been addressed by Bouhemadou *et al.* [5]. Cover *et al.* [6] also studied the elastic properties of this compund. Electronic properties have been studied by Music *et al.* [7]. Though elastic and electronic properties have been studied, the vibrational, thermodynamic and optical properties have not been theoretically explored so far. This paper attempts to bridge this gap through DFT based *ab-initio* calculations.

The thermodynamic properties of a compound are extremely important in solid state science and are considered as key factors in designing functional materials to be used under high temperature and high pressure conditions. Optical parameters, on the other hand, provide with the information about the electronic response of the material subject to incident electromagnetic radiation. Optical properties are intimately related to electronic band structure and topology of the incipient Fermi surface [8]. Therefore, an investigation of these properties is desirable both from the point of view of fundamental physics and potential industrial applications.

In the present work, we have aimed to add novel theoretical results to the existing lierature on the physical properties of $Sc_2AlC$ phase by using the first-principles method. We have especially focused on the vibrational, thermodynamic and optical properties. The rest of the paper is organized as follows. Section 2 describes the computational procedure in brief. Theoretical results of analysis are presented and discussed in detail in section 3. Major conclusions drawn from the theoretical findings can be found in section 4.

## 2. Computational methodology

The CASTEP (Cambridge serial total energy package) code [9] has been used to calculate the structural, vibrational, thermodynamic and optical pretties of $Sc_2AlC$. Ab-initio calculations use the plane wave pseudopotential approach based on the density functional theory (DFT)

[10]. The crystal parameters are obtained via geometry optimization which is performed through minimizing the total energy and internal forces by using the Broyden-Fletcher-Goldfarb-Shanno (BFGS) minimization technique [9]. During computations, the exchange-correlation is treated within the GGA (Generalized Gradient Approximation) PBE (Perdew-Burke-Ernzerhof) functional [11]. To sample the first Brillouin zone, a k-point grid of 9 × 9 × 2 mesh according to Monkhorst-Pack scheme [12] is employed for all calculations with a spacing of 0.02 Å$^{-1}$. The convergence of the planewave expansion is done with a kinetic energy cutoff of 500 eV. Excellent convergence is guaranteed by testing the Brillouin zone sampling and the kinetic energy cutoff which employs the tolerance for self-consistent field, energy, maximum force, maximum displacement, and maximum stress as $5.0\times10^{-7}$ eV/atom, $5.0\times10^{-6}$ eV/atom, 0.01 eV/Å, $5.0\times10^{-4}$ Å, and 0.02 GPa, respectively. Phonon dispersion was obtained using the DFPT linear-response method [13]. Quasi-harmonic approximation is used to obtain the thermodynamic properties from the phonon dispersion curve and phonon density of states.

## 3. Results and discussion

### 3.1. Structural properties

Sc$_2$AlC compound is known to crystallize under ambient conditions in the Cr$_2$AlC crystal structure, with space group *P63/mmc* (No. 194). The compound has eight atoms in each unit cell and the unit cell contains two formula units. The positions of atoms in Sc$_2$AlC are as follows: C atoms are placed at the positions (0, 0, 0), the Al atoms are at (1/3, 2/3, 3/4) and the Sc atoms are at (1/3, 2/3, $z_M$) [14]. The lattice parameters *a, c* and $z_M$ are used to define the crystal structure where *a* and *c* are lattice constants and $z_M$ is internal structural parameter. The optimized unit cell is shown in Fig. 1. The optimized values of structural parameters of Sc$_2$AlC are given in Table 1. Our results are in good agreement with the theoretical results [5, 7].

**Table 1**
Optimized lattice parameters (*a* and *c*, in Å), hexagonal ratio *c/a*, internal parameter $z_M$, unit cell volume *V* (Å$^3$) for MAX phases Sc$_2$AlC.

| Phases | a | c | c/a | $z_M$ | V | Ref |
|---|---|---|---|---|---|---|
| Sc$_2$AlC | 3.290 | 15.106 | 4.591 | 0.0821 | 141.600 | This |
| | 3.2275 | 14.8729 | 4.6081 | 0.0824 | 134.167* | [5] LDA |
| | 3.280 | 15.3734 | 4.687 | | 143.230* | [7] GGA |

*Calculated using $V = 0.866 a^2 c$

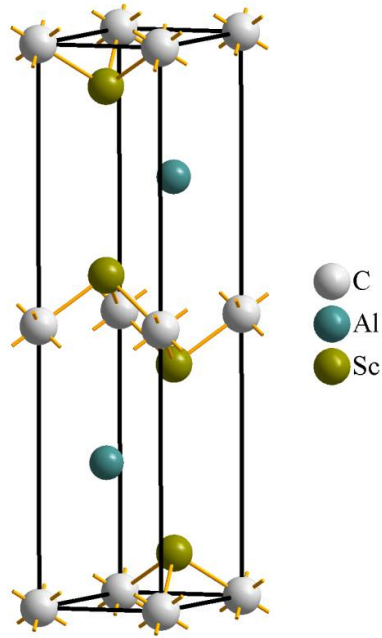

Fig. 1. The unit cell of Sc$_2$AlC MAX nanolaminate.

*3.2 Vibrational properties*

Fig. 2 presents the ground state (ambient) phonon dispersion curve and phonon density of states (PHDOS) along the high-symmetry directions of the crystal Brillouin zone. There are no experimental or theoretical data for available at this moment; therefore, a comparison is not possible at this time. The corresponding frequency for longitudinal optical (LO) and transverse optical (TO) modes at the zone centre are 17.3 THz and 12.2 THz, respectively.

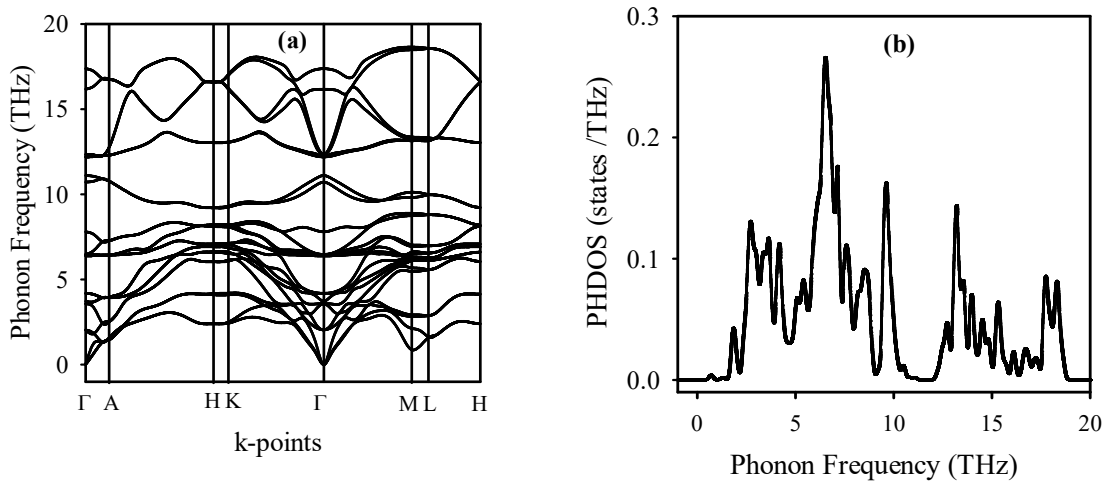

Fig. 2 (a) Phonon dispersion curve and (b) Phonon density of states of Sc$_2$AlC.

The highest point for LO is not located at zone centre but at the high symmetry point of M. The separation between LO and TO at the zone centre is 5.2 THz. A compound said to be dynamically stable if the phonon frequencies for all the wave vectors are positive. A compound is treated as dynamically unstable, if there is any imaginary phonon frequency at any wave vector. Since, in our present case all the frequencies are positive, therefore the phase under consideration is dynamically stable. Fig. 2(a) also depicts that there is a clear gap between acoustic branch and optical branch in the whole BZ as suggested by the phonon dispersion curves and PHDOS in Fig. 2(b).

*3.3 Thermodynamic properties*

A complete description of a system's equilibrium is contained in its thermodynamic potentials. We have obtained the thermodynamical potential functions such as Helmholtz free energy $F$, internal energy $E$, entropy $S$ and specific heat $C_v$ of $Sc_2AlC$ at zero pressure using the calculated phonon density of states emplying quasi-harmonic approximation [15]. The following equations have been used to calculate the $F$, $E$, $S$ and $C_v$ [16]:

$$F = 3nNk_BT \int_0^{\omega_{max}} \ln\left\{2\sinh\left(\frac{\hbar\omega}{2k_BT}\right)\right\} g(\omega)d\omega \quad \text{...............} (1)$$

$$E = 3nN\frac{\hbar}{2} \int_0^{\omega_{max}} \omega \coth\left(\frac{\hbar\omega}{2k_BT}\right) g(\omega)d\omega \quad \text{...............} (2)$$

$$S = 3nNk_B \int_0^{\omega_{max}} \left[\frac{\hbar\omega}{2k_BT}\coth\left(\frac{\hbar\omega}{2k_BT}\right) - \ln\left\{2\sinh\left(\frac{\hbar\omega}{2k_BT}\right)\right\}\right] g(\omega)d\omega \text{......} (3)$$

$$C_V = 3nNk_B \int_0^{\omega_{max}} \left(\frac{\hbar\omega}{2k_BT}\right)^2 \operatorname{csch}^2\left(\frac{\hbar\omega}{2k_BT}\right) g(\omega)d\omega \quad \text{...............} (4)$$

where, $k_B$ is the Boltzmann constant, $n$ is the number of atoms per unit cell, $N$ is the Avogadro's number, $\omega_{max}$ is the cut-off phonon frequency, $\omega$ is the phonon frequency, and $g(\omega)$ is the normalized phonon density of states, giving $\int_0^{\omega_{max}} g(\omega)\, d\omega = 1$.

The calculated $F$, $E$, $S$ and $C_v$ are shown in Fig. 3(a-d) in the temperature range from 0 to 1000 K. Helmholtz free energy ($F$) of $Sc_2AlC$ is displayed in Fig. 3(a) in which the free energy gradually decreases with increasing temperature. The decreasing trend of free energy is very common and it becomes more negative during the course of any natural process. The degree of decrease in free energy is determined by the enropy (*S*) of any system. The entropy of a system is increases with increasing temperature since thermal agitation adds to disorder. This is shown in Fig. 3(c). In contrary to the free energy, the internal energy (*E*) shows an increasing trend with temperature as shown in Fig. 3(b).

The behavior of materials under different thermodynamical constraints can be explained in terms of the specific heat of a solid. It also defines how efficiently the material stores heat. The phonon contribution dominates $C_v$ as a function of temperature. Fig. 3 (d) shows that the specific heat $C_v$ of $Sc_2AlC$ follows the Debye model which is proportional to $T^3$, as expected [17]. This model correctly predicts the temperature dependence of the heat capacity at constant volume at low temperature. It is seen that up to 300 K, the heat capacity of $Sc_2AlC$ increases rapidly with increase in temperature. It is also found that the Dulong-Petit law is recovered at high temperatures [18].

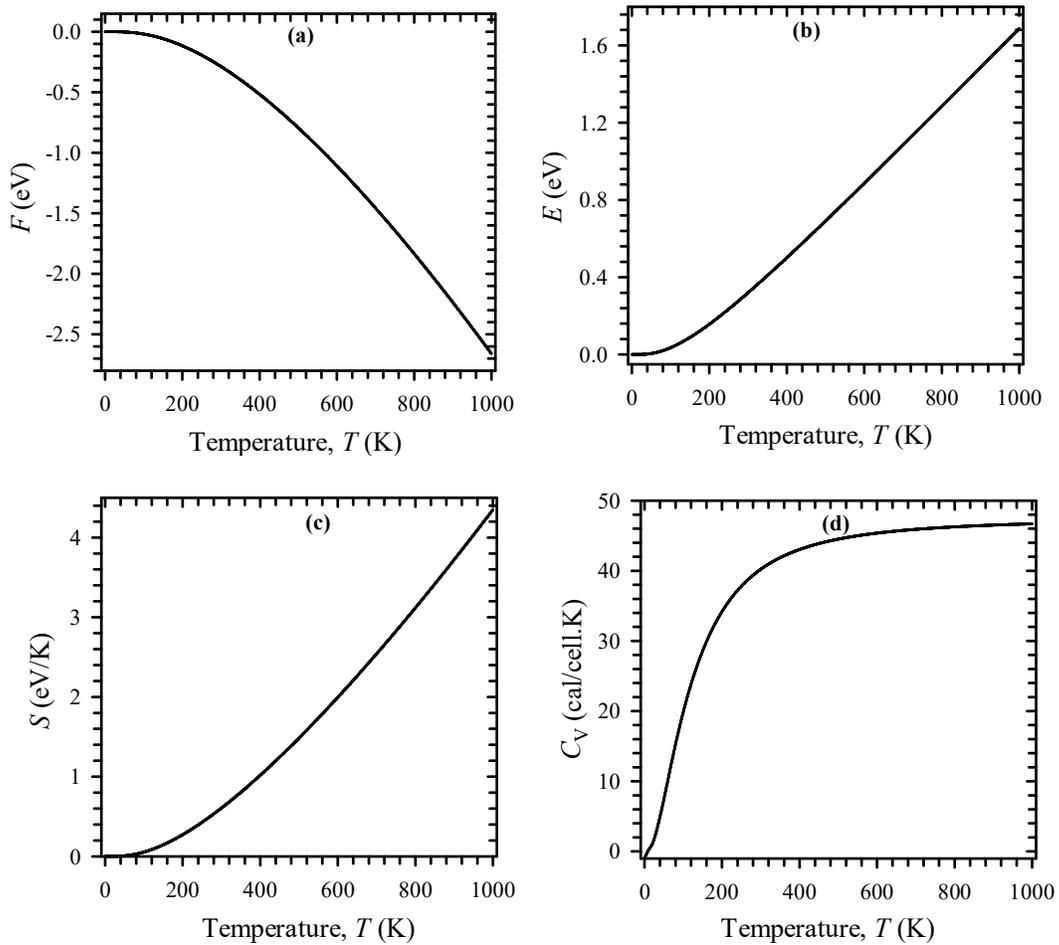

Fig. 3. Temperature dependence of the calculated thermodynamic parameters of $Sc_2AlC$ under zero pressure. (a) Helmholtz free energy, (b) Internal energy, (c) Entropy, and (d) Specific heat.

The Debye temperature, $\theta_D$, is an important parameter related to many thermophysical properties. In general, a high Debye temperature implies strong bonding among the atoms and a higher phonon contribution to thermal conductivity. The estimated $\theta_D$ from phonon spectrum is around 638 K. $\theta_D$ can also be calculated from various elastic constants. Following the procedure reported in [19], we have also calculated $\theta_D$ from elastic constants, which is 558 K.

*3.4. Optical properties*

When an electromagnetic radiation is incident on the materials, different materials behave in different way. The optical constants determine the overall response of the sample to the incident radiation. The complex dielectric functions, defined as $\varepsilon(\omega) = \varepsilon_1(\omega) + i\varepsilon_2(\omega)$ is one of the main optical characteristics of solids. The other optical constants can be extracted from this complex function. The imaginary part $\varepsilon_2(\omega)$ is calculated by CASTEP [20] numerically by a direct evaluation of the transition matrix elements between the occupied and unoccupied electronic states. The expression for the $\varepsilon_2(\omega)$ can be found elsewhere [21, 22]. The Kramers-Kronig (KK) relations are used to derive the real part $\varepsilon_1(\omega)$ of dielectric function from the calculated imaginary part. The other optical constants described in this section are derived from $\varepsilon_1(\omega)$ and $\varepsilon_2(\omega)$ using the equations given in Ref. 20. Information regarding optical constants are important in display technology.

The optical constants of $Sc_2AlC$ are shown in Fig. 4 (left and right panels) for the (100) polarization direction of the incident electric field. To smear out the Fermi level for effective *k*-points on the Fermi surface, we have used a 0.5 eV Gaussian smearing. A Drude term with an unscreened plasma energy of 3 eV and a damping term of width 0.05 eV have also been used.

Optical parameters give useful insight about the underlying electronic band structure. The electronic properties of crystalline material are mainly characterized by the imaginary part, $\varepsilon_2(\omega)$ of dielectric function, $\varepsilon(\omega)$, which is related to the photon absorption phenomenon. The peaks in $\varepsilon_2(\omega)$ are associated with electron excitations. For the compound under study, there is only one prominent peak at 2.85 eV (Fig. 4 (b)). The large negative value of $\varepsilon_1$ is also observed in Fig. 4 (a), which is a clear indication of Drude-like behavior seen in metals. The refractive index, *n*, is another technically important parameter for optoelectronic materials. The frequency dependent refractive index is shown in Fig. 4 (c). The extinction coeffient *k* is exhibited in Fig. 4 (d). The extinction coefficient measures the degree of attenuation of electromagnetic radiation inside the solid.

Fig. 4 (e) illustrates the behavior of the absorption coefficient spectra of $Sc_2AlC$. This reveals the metallic nature of the compound since the absorption coefficient is finite at zero energy. There is no discernible band gap and free carrier absorption dominates at low nergies.

The loss function, $L(\omega)$, is shown in Fig. 4 (f). The loss function measures the energy loss of an electron with high velocity passing through the material. This curve is characterized by a peak at an energy which gives the bulk plasma frequency, $\omega_P$, that occurs at the onset of $\varepsilon_2 < 1$ and $\varepsilon_1 = 0$.

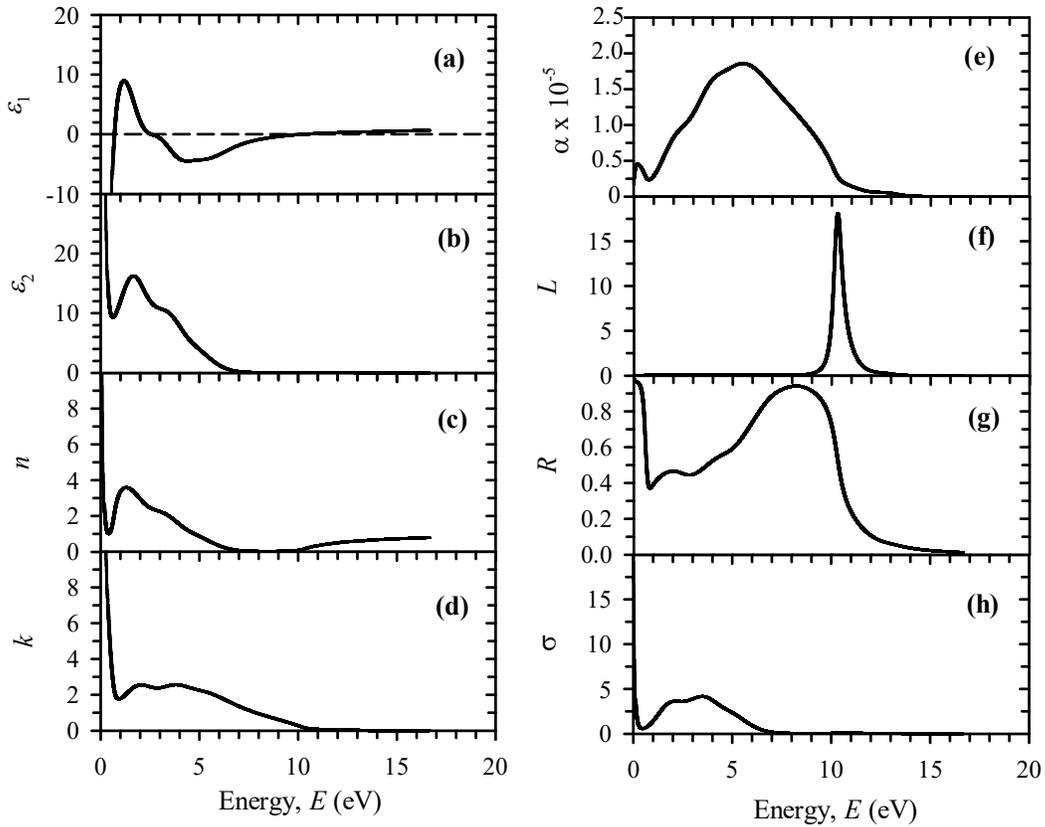

Fig. 4. The energy dependence of the calculated optical parameters of $Sc_2AlC$ – (a) Real part of dielectric constant (b) Imaginary part of the dielectric constant (c) Refractive index (d) Extinction coefficient (e) Absoption coefficient (f) Loss function (g) Reflectance (h) Optical conductivity.

From Fig. 4 (f) it is seen that the value of the effective plasma frequency $\omega_P$ is found to be ~ 10.3 eV. A metal becomes transparent, when the frequency of incident photon is greater than $\omega_P$. The reflectivity curve is shown in Fig. 4 (g). It is seen that the reflectance curve starts with a high value of ~ 0.90 – 0.98, decreases and then rises again to reach maximum value of ~ 0.80 – 0.90 between 6 and 10 eV. The large reflectivity at very low energies indicates that the dynamical conductivity is quite high for $Sc_2AlC$ in the low energy (frequency) region. Moreover, the peak of loss function is associated with the trailing edge of the reflection spectra, as theoretically expected.

Since the material under study has no band gap, the photoconductivity starts (with a high value) at zero photon energy as shown in Fig. 4 (h) reinforcing the fact $Sc_2AlC$ is metallic in nature.

## 4. Conclusions

The vibrational, thermodynamic and optical properties of $Sc_2AlC$ MAX compound have been investigated for the first time by means of DFT based first-principles method. Structural properties are compared and found to be in good agreement with available results. Phonon dispersion curve indicate that $Sc_2AlC$ is dynamically stable. Clear separation between acoustic and optical branches is seen. Thermodynamic properties are obtained from phonon density of states. The Debye temperature is relatively high. A high Debye temperature usually implies a high phonon thermal conductivity. The optical parameters, such as the real and imaginary parts of the dynamical susceptibilities, absorption coefficient, loss function, and photoconductivity spectra reveal the metallic nature of $Sc_2AlC$. The low energy optical conductivity is quite high. The reflectance spectrum of the compound under study shows that it might be used as a shielding material to avoid solar heating. We hope that the experimentalists will be encouraged to use the findings of this study to explore this material in greater details in near future.